\documentclass[dvipdfmx]{ptephy}

\usepackage{amsfonts}
\usepackage{amssymb}
\usepackage{amsmath}

\usepackage{ulem}
\usepackage{color}
\newcommand{\Slash}[1]{{\ooalign{\hfil/\hfil\crcr$#1$}}}

\def\Slash#1{\not\!\!#1}

\begin{document}

\title{
Analytical formulae of the Polyakov and the Wilson loops 
with Dirac eigenmodes in lattice QCD
}

\author{
  \name{\fname{Hideo}  \surname{Suganuma}}{1},
  \name{\fname{Takahiro M.} \surname{Doi}}{1}
  \name{\fname{Takumi} \surname{Iritani}}{2},
}

\address{
  \affil{1}
  {Department of Physics, Graduate School of Science,
   Kyoto University, 
   Kitashirakawa-oiwake, Sakyo, Kyoto 606-8502, Japan \\ 
   \email{suganuma@scphys.kyoto-u.ac.jp}\\
   \email{doi@ruby.scphys.kyoto-u.ac.jp}}
  \affil{2}
  {Yukawa Institute for Theoretical Physics (YITP), 
   Kyoto University, 
   Kitashirakawa-oiwake, Sakyo, Kyoto 606-8502, Japan \\
   \email{iritani@yukawa.kyoto-u.ac.jp}}
}

\begin{abstract}
We derive an analytical gauge-invariant formula 
between the Polyakov loop $L_P$ and 
the Dirac eigenvalues $\lambda_n$ in QCD, i.e., 
$L_P \propto 
\sum_n \lambda_n^{N_t -1} \langle n|\hat U_4|n \rangle$, 
in ordinary periodic square lattice QCD 
with odd-number temporal size $N_t$. 
Here, $|n\rangle$ denotes the Dirac eigenstate, and $\hat U_4$ 
temporal link-variable operator.
This formula is a Dirac spectral representation of 
the Polyakov loop in terms of Dirac eigenmodes $|n\rangle$.
Because of the factor $\lambda_n^{N_t -1}$ in the Dirac spectral sum, 
this formula indicates negligibly small contribution of 
low-lying Dirac modes to the Polyakov loop 
in both confinement and deconfinement phases, 
while these modes are essential for chiral symmetry breaking. 
Next, we find a similar formula between the Wilson loop and Dirac modes 
on arbitrary square lattices, without restriction of odd-number size. 
This formula suggests a small contribution of low-lying Dirac modes 
to the string tension $\sigma$, or the confining force. 
These findings support 
no crucial role of low-lying Dirac modes for confinement, 
i.e., no direct one-to-one correspondence between 
confinement and chiral symmetry breaking in QCD, 
which seems to be natural because heavy quarks are also confined 
even without light quarks or the chiral symmetry. 
\end{abstract}

\subjectindex{B01, B02, B03, B64}

\maketitle
\section{Introduction}

Since quantum chromodynamics (QCD) was established 
as the fundamental theory of strong interaction \cite{N66,GWP73}, 
it has been an important problem in theoretical physics 
to clarify color confinement and 
spontaneous chiral-symmetry breaking \cite{NJL61}.
However, in spite of many and various studies, 
these two nonperturbative phenomena 
have not been well understood directly from QCD. 

Dynamical chiral-symmetry breaking in QCD is categorized 
as well-known spontaneous symmetry breaking, 
which widely appears in various phenomena in physics. 
The standard order parameter of chiral symmetry breaking 
is the quark condensate $\langle \bar qq \rangle$, 
and it is directly related to low-lying Dirac modes, 
as the Banks-Casher relation indicates \cite{BC80}.
Here, Dirac modes are eigenmodes of the Dirac operator 
$\Slash D$, which directly appears in the QCD Lagrangian.

In contrast to chiral symmetry breaking, 
color confinement is a quite unique phenomenon 
peculiar in QCD, and the quark confinement is characterized by 
the area law of the Wilson loop, i.e., non-zero string tension, 
or the zero Polyakov loop, i.e., infinite single-quark free energy. 

The Polyakov loop $L_P$ is 
one of the typical order parameters, and 
it relates to the single-quark free energy $E_q$ as 
$\langle L_P \rangle \propto e^{-E_q/T}$ at temperature $T$. 
The Polyakov loop is also an order parameter 
of spontaneous breaking of the $Z_{N_c}$ center symmetry 
in QCD \cite{R12}.

In addition to the study of each nonperturbative phenomenon, 
to clarify the relation between confinement 
and chiral symmetry breaking is one of 
the challenging important subjects in theoretical physics
\cite{SST95,M95,W95,G06BGH07,SWL08,S111213,GIS12,IS14,SDI13,DSI14},
and their relation is not yet clarified directly from QCD.

A strong correlation between confinement and chiral symmetry breaking 
has been suggested by 
almost coincidence between deconfinement and chiral-restoration temperatures 
\cite{R12,K02}, although slight difference of about 25MeV between them 
is pointed out in recent lattice QCD studies \cite{AFKS06}.
Their correlation has been also suggested 
in terms of QCD-monopoles \cite{SST95,M95,W95}, 
which topologically appear in QCD in the maximally Abelian gauge. 
By removing the monopoles from the QCD vacuum, 
confinement and chiral symmetry breaking 
are simultaneously lost \cite{SST95,M95,W95}, 
which indicates an important role of QCD-monopoles 
to both phenomena, and thus 
these two phenomena seem to be related via the monopole.

As another type of pioneering study, Gattringer and Bruckmann {\it et al.} 
showed that the Polyakov loop can be analytically expressed with 
the Dirac eigenvalues under the temporally twisted boundary condition 
for temporal link-variables \cite{G06BGH07}.
Although temporal (nontwisted) periodic boundary condition 
is physically required for link-variables in real QCD at finite temperature, 
such an analytical formula would be useful 
to consider the relation between confinement and chiral symmetry breaking.

In a series of our recent studies \cite{S111213,GIS12,IS14}, 
we have numerically investigated the Wilson loop and the Polyakov loop 
in terms of the ``Dirac-mode expansion'', 
and have found that quark confinement properties are almost kept 
even in the absence of low-lying Dirac modes. 
(Also, ``hadrons'' appear without low-lying Dirac modes 
 \cite{LS11}, suggesting survival of confinement.)
Note that the Dirac-mode expansion is just 
a mathematical expansion by eigenmodes $|n \rangle$ of 
the Dirac operator $\Slash D=\gamma_\mu D_\mu$, 
using the completeness of $\sum_n|n \rangle \langle n|=1$. 
In general, instead of $\Slash D$, 
one can consider any (anti)hermitian operator, e.g., $D^2=D_\mu D_\mu$, 
and the expansion in terms of its eigenmodes \cite{BI05}. 
To investigate chiral symmetry breaking, however, 
it is appropriate to consider $\Slash D$ 
and the expansion by its eigenmodes.

In this paper, we derive analytical formulae of 
the Polyakov and the Wilson loops with the Dirac modes 
in the lattice QCD formalism \cite{SDI13,DSI14}, 
and discuss the relation between confinement and chiral symmetry breaking.

The organization of this paper is as follows. 
In Sect.~2, we briefly review the lattice QCD formalism for 
the Dirac operator, Dirac eigenvalues and Dirac modes. 
In Sect.~3, we derive an analytical formula between the Polyakov loop 
and the Dirac modes in lattice QCD where the temporal size is odd-number.
In Sect.~4, we investigate the properties of the obtained formula, 
and discuss the contribution from the low-lying Dirac modes 
to the Polyakov loop.
In Sect.~5, we consider the relation between the Wilson loop and Dirac modes
on arbitrary square lattices, without restriction of odd-number size.
Section~6 will be devoted to the summary.

\section{Lattice QCD formalism}

To begin with, we state the setup condition of 
lattice QCD formalism adopted in this study.
We use an ordinary square lattice with spacing $a$ and 
size $N_s^3 \times N_t$.
The normal nontwisted periodic boundary condition is used 
for the link-variable $U_\mu(s)={\rm e}^{iagA_\mu(s)}$ 
in the temporal direction, 
with the gluon field $A_\mu(s)$, the gauge coupling $g$ and the site $s$.
This temporal periodicity is physically required at finite temperature.
In this paper, we take SU($N_c$) with $N_c$ being the color number 
as the gauge group of the theory. 
However, arbitrary gauge group $G$ can be taken 
for most arguments in the following. 

\subsection{Lattice QCD formalism and anatomy of gauge ensemble}

In the Euclidean lattice formalism, 
the QCD generating functional is expressed with 
the QCD action $S_{\rm QCD}$ as 
\begin{eqnarray}
Z_{\rm QCD}=
\int D{\bar q} Dq DU e^{-S_{\rm QCD}} 
\int D{\bar q} Dq DU e^{-\{S_{\rm gauge}[U]+{\bar q}K[U]q\}} 
 =\int DU e^{-S_{\rm gauge}[U]}{\rm det} K[U],
\label{eq:QCDgf}
\end{eqnarray}
where $S_{\rm gauge}[U]$ denotes the lattice gauge action 
and $K[U]$ a fermionic kernel.
In this study, one can freely choose any type of lattice fermions such as 
the Wilson fermion, the Kogut-Susskind fermion, the overlap fermion, 
and so on \cite{R12}.
As importance sampling for the generating function $Z$, 
one can generate gauge configurations $\{U_k\}_{k=1,2,3, ..., N}$ 
using Monte Carlo simulations.
The expectation value of any operator $O[U]$ is given by 
the gauge ensemble average as 
\begin{eqnarray}
\langle O[U] \rangle 
=\frac{1}{{Z}_{\rm QCD}} 
\int DU e^{-S_{\rm gauge}[U]}{\rm det} K[U] \cdot O[U] 
={\rm lim}_{N \rightarrow \infty} \frac{1}{N}\sum_{k=1}^N O[U_k].
\end{eqnarray}
In this study, we consider some analytical relations 
between Dirac modes and confinement properties for 
the gauge configurations $\{U_k\}_{k=1,2,3, ..., N}$, 
generated in full QCD or quenched QCD with setting ${\rm det} K[U]=1$.


In this paper, we perform ``anatomy" for the nonperturbative QCD vacuum, 
i.e., the gauge ensemble which is in principle 
generated for the QCD generating functional $Z_{\rm QCD}$.
In our approach, 
we do not change the QCD action $S_{\rm QCD}$ or $Z_{\rm QCD}$,
but analyze the QCD vacuum for $Z_{\rm QCD}$ in terms of the Dirac modes. 
This approach is similar to our previous works 
\cite{S111213,GIS12,IS14} and the works by Lang~et~al. \cite{LS11}, 
where low-lying Dirac modes are removed from the lattice QCD configurations, 
after their numerical generation 
in standard Monte Carlo simulations for $Z_{\rm QCD}$. 
In these studies, $Z_{\rm QCD}$ is not changed at all. 
Our approach is also similar to that of Abelian dominance \cite{SS14} 
and monopole dominance \cite{SNW94} for the argument of quark confinement 
in the MA gauge. 
After generating QCD configuration in the MA gauge, 
off-diagonal gluons or monopoles are removed from the QCD vacuum, 
and confinement properties are investigated in the processed QCD vacuum. 
In these studies, $Z_{\rm QCD}$ is also unchanged. 
(If off-diagonal gluons are removed from $Z_{\rm QCD}$ at the action level, 
the system becomes QED, which is no more meaningful for the study of QCD.) 
In fact, the main interest is the QCD vacuum, 
and it has been investigated from the viewpoint of some relevant modes, 
such as low-lying Dirac modes, monopoles and so on.
In this work, we analyze the lattice gauge ensemble, 
generated for $Z_{\rm QCD}$, in terms of the Dirac modes.

\subsection{Dirac operator, Dirac eigenvalues and Dirac modes in lattice QCD}

Here, we mathematically define 
the Dirac operator $\Slash{D}$, Dirac eigenvalues $\lambda_n$, 
and Dirac eigenmodes $|n \rangle$ in lattice QCD.

In lattice QCD, the Dirac operator 
$\Slash D = \gamma_\mu D_\mu$ is expressed with 
$U_\mu(s)={\rm e}^{iagA_\mu(s)}$.
In our study, we take the lattice Dirac operator of 
\begin{eqnarray}
 \Slash{D}_{s,s'} 
 \equiv \frac{1}{2a} \sum_{\mu=1}^4 \gamma_\mu 
\left[ U_\mu(s) \delta_{s+\hat{\mu},s'}
 - U_{-\mu}(s) \delta_{s-\hat{\mu},s'} \right],
\label{eq:latticedirac}
\end{eqnarray}
where $\hat\mu$ is the unit vector in $\mu$-direction in the lattice unit, 
and $U_{-\mu}(s)\equiv U^\dagger_\mu(s-\hat \mu)$.
Adopting hermitian $\gamma$-matrices as $\gamma_\mu^\dagger=\gamma_\mu$, 
the Dirac operator $\Slash D$ is anti-hermitian and satisfies 
$\Slash D_{s',s}^\dagger=-\Slash D_{s,s'}$.
(Note that the Dirac operator $\Slash{D}_{s,s'}$ defined here 
is not identical with the fermionc kernel $K_{s,s'}[U]$ in Sec.~2.1.
The relation between $\Slash D$ and $K[U]$ will be discussed in Sec.~2.4.)

We introduce the normalized Dirac eigen-state $|n \rangle$ as 
\begin{eqnarray}
\Slash D |n\rangle =i\lambda_n |n \rangle, \qquad
\langle m|n\rangle=\delta_{mn}, 
\end{eqnarray}
with the Dirac eigenvalue $i\lambda_n$ ($\lambda_n \in {\bf R}$).
Because of $\{\gamma_5,\Slash D\}=0$, the state 
$\gamma_5 |n\rangle$ is also an eigen-state of $\Slash D$ with the 
eigenvalue $-i\lambda_n$. 
Here, the Dirac eigen-state $|n \rangle$ 
satisfies the completeness of 
\begin{eqnarray}
\sum_n |n \rangle \langle n|=1.
\end{eqnarray}
For the Dirac eigenfunction $\psi_n(s)\equiv\langle s|n \rangle$, 
the Dirac eigenvalue equation 
$\Slash D \psi_n(s)=i\lambda_n \psi_n(s)$
is expressed by  
\begin{eqnarray}
\sum_{s'}\Slash D_{s,s'} \psi_n(s')=i\lambda_n \psi_n(s)
\end{eqnarray}
in lattice QCD, and its explicit form is written by 
\begin{eqnarray}
\frac{1}{2a} \sum_{\mu=1}^4 \gamma_\mu
[U_\mu(s)\psi_n(s+\hat \mu)-U_{-\mu}(s)\psi_n(s-\hat \mu)] 
=i\lambda_n \psi_n(s).
\end{eqnarray}
The Dirac eigenfunction $\psi_n(s)$ can be 
numerically obtained in lattice QCD, besides a phase factor. 
By the gauge transformation of 
$U_\mu(s) \rightarrow V(s) U_\mu(s) V^\dagger (s+\hat\mu)$, 
$\psi_n(s)$ is gauge-transformed as 
\begin{eqnarray}
\psi_n(s)\rightarrow V(s) \psi_n(s),
\label{eq:GTDwf}
\end{eqnarray}
which is the same as that of the quark field, although, 
to be strict, there can appear an irrelevant $n$-dependent 
global phase factor $e^{i\varphi_n[V]}$, 
according to arbitrariness of the phase in 
the basis $|n \rangle$
\cite{GIS12}.

Note that the spectral density $\rho(\lambda)$ 
of the Dirac operator $\Slash D$ relates to chiral symmetry breaking 
in continuum QCD.
For example, from Banks-Casher's relation \cite{BC80}, 
the zero-eigenvalue density $\rho(0)$ leads to 
$\langle\bar qq \rangle$ as 
\begin{eqnarray}
\langle \bar qq \rangle&=&-\lim_{m \to 0} \lim_{V_{\rm phys} \to \infty} 
\pi\rho(0), \\
\rho(\lambda)&\equiv& 
\frac{1}{V_{\rm phys}}\sum_{n}\langle \delta(\lambda-\lambda_n)\rangle,
\end{eqnarray}
with space-time volume $V_{\rm phys}$. 
(In lattice QCD, the use of the Dirac operator $\Slash D$ 
in Eq.(\ref{eq:latticedirac}) accompanies 
an overall degeneracy factor $2^4$, which will be discussed in Sec.2.4.) 
In any case, the low-lying Dirac modes can be regarded as the essential modes 
responsible to spontaneous chiral-symmetry breaking in QCD.

\subsection{Operator formalism in lattice QCD}

Now, we present the operator formalism 
in lattice QCD \cite{S111213,GIS12,IS14}. 
To begin with, we introduce the link-variable operator 
$\hat U_\mu$ ($\mu=\pm 1,... , \pm 4$) 
defined by the matrix element of 
\begin{eqnarray}
\langle s |\hat U_{\pm \mu}|s' \rangle 
=U_{\pm \mu}(s)\delta_{s\pm \hat \mu,s'}.
\label{eq:LVO}
\end{eqnarray}
Because of $U_{-\mu}(s)=U_\mu^\dagger(s-\hat \mu)$, 
$\hat U_{\pm \mu}$ are hermite conjugate each other and satisfy 
\begin{eqnarray}
\hat U_{-\mu}=\hat U_{\mu}^\dagger.
\end{eqnarray}
With the link-variable operator $\hat U_{\pm \mu}$, 
the covariant derivative is written as 
\begin{eqnarray}
\hat D_\mu=\frac{1}{2a}(\hat U_\mu-\hat U_{-\mu}), 
\end{eqnarray}
and the Dirac operator defined by Eq.(\ref{eq:latticedirac}) 
is simply expressed as 
\begin{eqnarray}
\Slash{\hat D}
=\frac{1}{2a}\sum_{\mu=1}^{4} \gamma_\mu (\hat U_\mu-\hat U_{-\mu}).
\label{eq:Dop}
\end{eqnarray}
Both $\Slash{\hat D}$ and $\hat D_\mu$ are anti-hermite operators.
The Dirac-mode matrix element of the link-variable operator 
$\hat U_{\mu}$ ($\mu=\pm 1,..., \pm 4$) can be expressed with $\psi_n(s)$:
\begin{eqnarray}
\langle m|\hat U_{\mu}|n \rangle
=\sum_s\langle m|s \rangle 
\langle s|\hat U_{\mu}|s+\hat \mu \rangle \langle s+\hat \mu|n\rangle
=\sum_s \psi_m^\dagger(s) U_\mu(s)\psi_n(s+\hat \mu).
\end{eqnarray}
Note that the matrix element is gauge invariant, 
apart from an irrelevant phase factor.
Actually, using the gauge transformation (\ref{eq:GTDwf}), we find 
the gauge transformation of the matrix element as \cite{GIS12}
\begin{eqnarray}
\langle m|\hat U_\mu|n \rangle
&=&\sum_s \psi^\dagger_m(s)U_\mu(s)\psi_n(s+\hat\mu) \cr
&\rightarrow&
\sum_s\psi^\dagger_m(s)V^\dagger(s)\cdot V(s)U_\mu(s)V^\dagger(s+\hat \mu) 
\cdot V(s+\hat \mu)\psi_n(s+\hat \mu) \cr
&=&\sum_s\psi_m^\dagger(s)U_\mu(s)\psi_n(s+\hat \mu)
=\langle m|\hat U_\mu|n\rangle.
\end{eqnarray}
To be strict, there appears an $n$-dependent global phase factor, 
corresponding to the arbitrariness of the phase in the basis 
$|n \rangle$. However, this phase factor cancels 
as $e^{i\varphi_n} e^{-i\varphi_n}=1$ 
between $|n \rangle$ and $\langle n |$, and does not appear 
for physical quantities such as the Wilson loop and the Polyakov loop 
\cite{GIS12}.


\subsection{Relation between Dirac operator $\Slash D$ and 
fermionic kernel $K$}

In this subsection, we discuss 
the relation between the Dirac operator 
$\Slash D$ defined in Eq.(\ref{eq:latticedirac}) or (\ref{eq:Dop})
and the fermionic kernel $K[U]$ in lattice QCD in Eq.(\ref{eq:QCDgf}). 

In lattice QCD, the simple Dirac operator $\Slash D$ has $2^D$ degeneracy 
with $D=4$ being the number of space-time dimension \cite{R12}.
In the fermionic kernel $K[U]$, the doubler contribution 
is effectively removed in some way. 
For a typical example of the Wilson fermion, 
a large extra energy of $O(1/a)$ is added only for doublers, 
which makes the doublers inactive in the low-energy region. 
Then, 16 degenerate low-lying Dirac modes of $\Slash D$ 
correspond to one low-lying mode and 15 doubler modes 
in terms of the fermionic kernel $K[U]$.

In fact, each low-lying mode of $K[U]$ is expected to 
have a large overlap with an eigenmode of $\Slash D$, 
but the chiral property is largely different 
between $\Slash D$ and $K[U]$. 
Actually, if $\Slash D$ is misleadingly used 
instead of $K[U]$ in Eq.(\ref{eq:QCDgf}), 
the theoretical structure may be largely changed 
due to many flavors of 16 species \cite{H09}. 
In addition, the axial anomaly is totally different, 
since it is not broken in the Dirac operator $\Slash D$ 
due to the cancellation from the doublers \cite{R12}. 

Here, we denote by $|\nu \rangle \rangle_K$ 
the normalized mode of the fermionic kernel $K[U]$, 
to distinguish it from the Dirac mode $|n \rangle$.
Because of $\sum_n |n \rangle \langle n|=1$, one finds the identity,
\begin{eqnarray}
|\nu \rangle \rangle_K=\sum_n |n \rangle \langle n| \nu \rangle \rangle_K.
\end{eqnarray}

In this paper, we assume that 
each low-lying mode of the fermionic kernel $K[U]$ 
is mainly expressed with the low-lying Dirac modes of $\Slash D$.
This assumption does not mean one-to-one correspondence 
between a low-lying Dirac mode and a low-lying mode of $K[U]$, 
but saturation of each low-lying mode of $K[U]$ by 
low-lying Dirac modes of some range.
In fact, for each low-lying mode $|\nu \rangle \rangle_K$, we assume 
\begin{eqnarray}
|\nu \rangle \rangle_K
\simeq \sum_{{\rm low-lying}~n} |n \rangle \langle n| \nu \rangle \rangle_K,
\end{eqnarray}
which means 
\begin{eqnarray}
\sum_{{\rm low-lying}~n} |\langle n| \nu \rangle \rangle_K|^2 \simeq 1.
\label{eq:assumption}
\end{eqnarray}
Here, $\sum_{{\rm low-lying}~n}$ denotes the sum over low-lying Dirac modes. 
This assumption would be natural, however, 
it is desired to examine (\ref{eq:assumption}) 
quantitatively in lattice QCD \cite{DS15}.
From this assumption, 
if the low-lying Dirac modes of $\Slash D$ are removed, 
the low-lying modes $|\nu \rangle \rangle_K$ 
of the fermionic kernel $K[U]$ are also removed approximately. 
In fact, this assumption links the Dirac-mode expansion 
to the low-lying modes of $K[U]$, 
which is more directly connected to chiral symmetry breaking.

\subsection{Dirac operator and Polyakov loop in finite temperature QCD}

In this subsection, we investigate 
the Dirac operator and the Polyakov loop in finite temperature QCD.
In the imaginary-time formalism, 
the finite-temperature system requires 
periodicity for bosons and anti-periodicity for fermions 
in Euclidean temporal direction \cite{R12}.
Here, we consider such a temporally-(anti)periodic lattice 
with the temporal size $N_t$, 
which corresponds to the temperature $T=1/(N_ta)$.
In this thermal system, 
any fermion field $\psi(s)$ obeys 
\begin{eqnarray}
\psi(s+N_t\hat t)=-\psi(s),
\end{eqnarray}
with $\hat t=\hat 4$,
and the temporal anti-periodicity of quarks 
also reflects in the Dirac operator $\hat{\Slash D}$. 
In fact, the temporal structure of the matrix $\hat D_4$ which acts on quarks
is expressed as \cite{BBW13}
\begin{eqnarray}
\hat D_4 =
\frac{1}{2a}
\begin{pmatrix}
0 & U_4(1) & 0 & \cdots & 0 & U_4^\dagger({N_t}) \\
-U_4^\dagger(1) & 0 & U_4(2) & \cdots & 0 & 0 \\
0 & -U_4^\dagger(2) & 0 & \cdots & 0 & 0 \\
\vdots & \vdots & \vdots & \ddots & \vdots &\vdots \\
0 & 0 & 0 & \cdots & 0 & U_4({N_t-1}) \\
-U_4({N_t}) & 0 & 0 & \cdots & -U_4^\dagger({N_t-1}) & 0
\end{pmatrix},
\label{eq:explicit-dirac}
\end{eqnarray}
where $U_4(t) \equiv U_4({\bf s},t)$ ($t=1,2, ..., N_t$) 
is an abbreviation of the temporal link-variable.
The additional minus sign in front 
of $U_4({N_t})$ and $U_4^\dagger({N_t})$ 
reflects the anti-periodicity of quarks 
in the temporal direction.

For the thermal system, 
the link-variable operator $\hat U_{\pm \mu}$ is 
basically defined by the matrix element (\ref{eq:LVO}).
However, taking account of the temporal anti-periodicity 
in $\hat D_4$ acting on quarks, it is convenient 
to add a minus sign to the matrix element of 
the temporal link-variable operator $\hat U_{\pm 4}$ 
at the temporal boundary of $t=N_t(=0)$:
\begin{eqnarray}
\langle {\bf s}, N_t|\hat U_4| {\bf s}, 1 \rangle 
=-U_4({\bf s}, N_t),
\quad
\langle {\bf s}, 1|\hat U_{-4}| {\bf s}, N_t \rangle 
=-U_{-4}({\bf s}, 1)=-U_4^\dagger({\bf s}, N_t). 
\label{eq:LVthermal}
\end{eqnarray}
Here, $\hat U_{-\mu}=\hat U_\mu^\dagger$ is satisfied.
%
%
For thermal QCD, by using this definition of 
the link-variable operator $\hat U_{\pm \mu}$, 
the Dirac operator and the covariant derivative are also simply expressed as 
\begin{eqnarray}
\Slash{\hat D}
=\frac{1}{2a}\sum_{\mu=1}^{4} \gamma_\mu (\hat U_\mu-\hat U_{-\mu}), \quad 
\hat D_\mu = \frac{1}{2a}(\hat U_\mu-\hat U_{-\mu}),
\end{eqnarray}
which are consistent with Eq.(\ref{eq:explicit-dirac}).

The Polyakov loop $L_P$ 
is also simply written as the functional trace of 
$\hat U_4^{N_t}$,
\begin{eqnarray}
L_P = -\frac{1}{N_c V} {\rm Tr}_c \{\hat U_4^{N_t}\}
=\frac{1}{N_c V}
\sum_s {\rm tr}_c \{\prod_{n=0}^{N_t-1} U_4(s+n\hat t)\},
\label{eq:PL}
\end{eqnarray}
with the four-dimensional lattice volume $V \equiv N_s^3 \times N_t$.
Here, ``${\rm Tr}_c$'' denotes the functional trace 
of ${\rm Tr}_c \equiv \sum_s {\rm tr}_c$ with 
the trace ${\rm tr}_c$ over color index.
The minus sign stems from the additional minus on $U_4({\bf s}, N_t)$ 
in Eq.(\ref{eq:LVthermal}).

\section{Analytical formula between Polyakov loop and Dirac modes 
in lattice QCD with odd temporal size}

Now, we consider lattice QCD 
with odd-number temporal lattice size $N_t$, as shown in Fig.1. 
Here, we use an ordinary square lattice 
with the normal nontwisted periodic boundary condition 
for the link-variable in the temporal direction. 
(Of course, this temporal periodicity is 
physically required at finite temperature.)
The spatial lattice size $N_s$ is taken to be larger than $N_t$, 
i.e., $N_s > N_t$. 
Note that, in the continuum limit of $a \rightarrow 0$ and 
$N_t \rightarrow \infty$, 
any number of large $N_t$ gives the same physical result.
Then, in principle, it is no problem to use the odd-number lattice.

\begin{figure}[h]
\begin{center}
\includegraphics[scale=0.75]{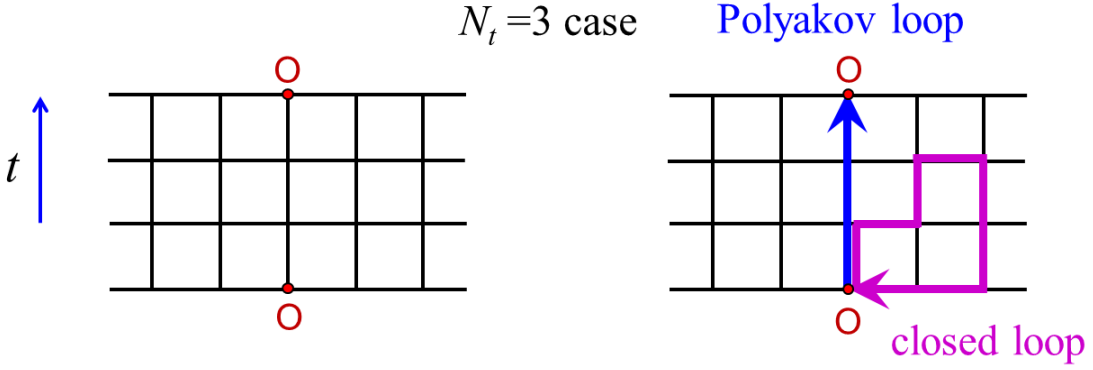}
\caption{
An example of the lattice with odd-number temporal size ($N_t=3$ case).
Only gauge-invariant quantities such as 
closed loops and the Polyakov loop survive or do not vanish in QCD, 
after taking the expectation value, i.e., the gauge-configuration average.
Geometrically, closed loops have even-number links on the square lattice.
}
\end{center}
\end{figure}

In general, only gauge-invariant quantities 
such as closed loops and the Polyakov loop 
survive in QCD, according to the Elitzur theorem \cite{R12}.
All the non-closed lines are gauge-variant 
and their expectation values are zero.
Note here that any closed loop, except for the Polyakov loop, 
needs even-number link-variables on the square lattice, 
as shown in Fig.1.

Note also that, from the definition of the link-variable operator 
$\hat U_\mu$ ($\mu \in \{\pm 1, ... , \pm 4\}$) in Eq.(\ref{eq:LVO}), 
the functional trace of the product of $\hat U_{\mu_k}$ 
along any non-closed trajectory is zero, i.e., 
\begin{eqnarray}
&&{\rm Tr}_{c} (\hat{U}_{\mu_1} \hat{U}_{\mu_2} \cdots \hat{U}_{\mu_N})
={\rm tr}_c 
\sum_s \langle s|\hat{U}_{\mu_1} \hat{U}_{\mu_2} \cdots \hat{U}_{\mu_N} |s 
\rangle \cr
&&=
{\rm tr}_c 
\sum_s 
U_{\mu_1}(s) 
U_{\mu_2}(s+\hat \mu_1) 
\cdots 
U_{\mu_N}(s+\sum_{k=1}^{N-1} \hat \mu_k) 
\langle s+\sum_{k=1}^N \hat \mu_k|s \rangle 
=0
\end{eqnarray}
for the non-closed trajectory with $\sum_{k=1}^N \hat \mu_k \ne 0$. 
(Here, $\hat \mu_k$ can take positive or negative direction 
as $\hat \mu_k \in \{\pm \hat 1, ..., \pm \hat 4 \}$, 
and any closed loop satisfies $\sum_{k=1}^N \hat \mu_k =0$.) 

In lattice QCD with odd-number temporal size $N_t$, 
we consider the functional trace of 
\begin{eqnarray}
I\equiv {\rm Tr}_{c,\gamma} (\hat{U}_4\hat{\Slash{D}}^{N_t-1}), 
\label{eq:FT}
\end{eqnarray}
where 
${\rm Tr}_{c,\gamma}\equiv \sum_s {\rm tr}_c 
{\rm tr}_\gamma$ includes 
${\rm tr}_c$ 
and the trace ${\rm tr}_\gamma$ over spinor index.
Its expectation value 
\begin{eqnarray}
 \langle I\rangle=\langle {\rm Tr}_{c,\gamma} (\hat{U}_4\hat{\Slash{D}}^{N_t-1})\rangle 
\label{eq:FTV}
\end{eqnarray}
is obtained as the gauge-configuration average in lattice QCD.
In the case of enough large volume $V$, one can expect 
$\langle O \rangle \simeq {\rm Tr}~O/{\rm Tr}~1$ 
for any operator $O$ at each gauge configuration.

From Eq.(\ref{eq:Dop}), 
$\hat U_4\!\Slash{\hat D}^{N_t-1}$ 
is expressed as a sum of products of $N_t$ link-variable operators, 
since the Dirac operator $\Slash{\hat D}$ 
includes one link-variable operator in each direction of $\pm \mu$.
Then, $\hat U_4\!\Slash{\hat D}^{N_t-1}$ 
includes many trajectories with the total length $N_t$ in the lattice unit 
on the square lattice, as shown in Fig.2.
Note that all the trajectories with the odd-number length $N_t$ 
cannot form a closed loop 
on the square lattice, and thus give gauge-variant contribution, 
except for the Polyakov loop.

\begin{figure}[h]
\begin{center}
\includegraphics[scale=0.75]{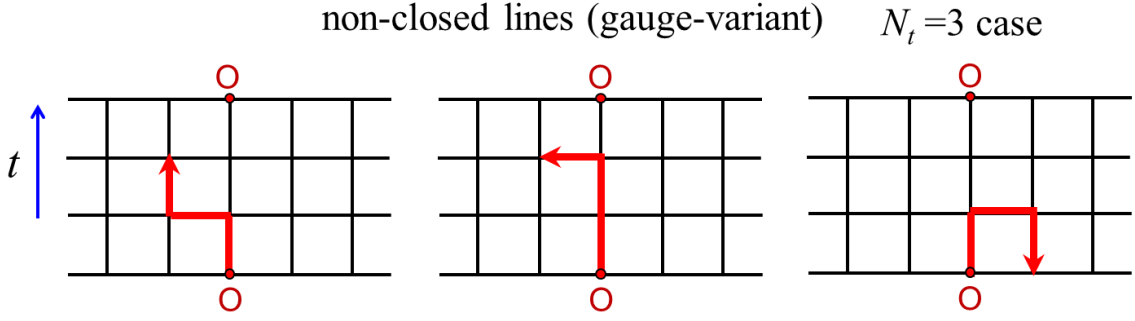}
\caption{
Partial examples of the trajectories stemming from 
$I \equiv {\rm Tr}_{c,\gamma}(\hat U_4\!\!\Slash{\hat D}^{N_t-1})$. 
For each trajectory, the total length is $N_t$, and 
the ``first step'' is positive temporal direction 
corresponding to $\hat U_4$.
All the trajectories with the odd-number length $N_t$ 
cannot form a closed loop on the square lattice, 
and therefore they are gauge-variant and give no contribution 
in $I$, except for the Polyakov loop.
}
\end{center}
\end{figure}

Therefore, among the trajectories stemming from 
${\rm Tr}_{c,\gamma}(\hat U_4\!\!\Slash{\hat D}^{N_t-1})$, 
all the non-loop trajectories are gauge-variant and give no contribution, 
according to the Elitzur theorem \cite{R12}.
Only the exception is the Polyakov loop. (See Figs.2 and 3.)
For each trajectory in $\hat U_4\!\!\Slash{\hat D}^{N_t-1}$, 
the first step is positive 
temporal direction corresponding to $\hat U_4$,
and hence ${\rm Tr}_{c,\gamma}(\hat U_4\!\!\Slash{\hat D}^{N_t-1})$ 
cannot include the anti-Polyakov loop $L_P^\dagger$.
Thus, in the functional trace 
$I
={\rm Tr}_{c,\gamma}(\hat U_4\!\!\Slash{\hat D}^{N_t-1})$, 
only the Polyakov-loop ingredient can survive 
as the gauge-invariant quantity, and 
$I$ is proportional to the Polyakov loop $L_P$.

\begin{figure}[h]
\begin{center}
\includegraphics[scale=0.5]{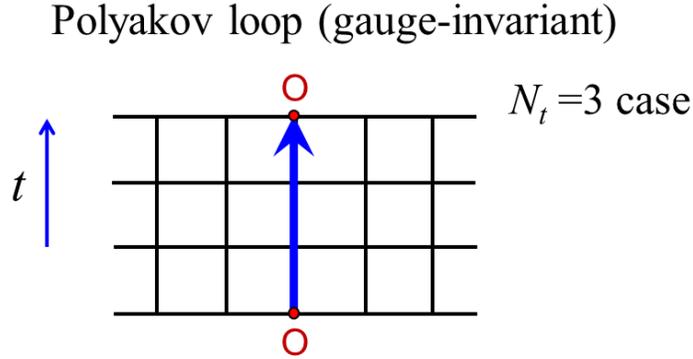}
\caption{
Among the trajectories stemming from 
${\rm Tr}_{c,\gamma}(\hat U_4\Slash{\hat D}^{N_t-1})$, 
only the Polyakov-loop ingredient can survive 
as a gauge-invariant quantity. 
Here, ${\rm Tr}_{c,\gamma}(\hat U_4\Slash{\hat D}^{N_t-1})$ 
does not include $L_P^\dagger$, 
because of the first factor $\hat U_4$.
}
\end{center}
\end{figure}

Actually, we can mathematically derive the following relation:
\begin{eqnarray}
I&=& {\rm Tr}_{c,\gamma} (\hat U_4 \hat{\Slash D}^{N_t-1}) 
= {\rm Tr}_{c,\gamma} \{\hat U_4 (\gamma_4 \hat D_4)^{N_t-1}\}
=4 {\rm Tr}_{c} (\hat U_4 \hat D_4^{N_t-1})\cr
&=&\frac{4}{(2a)^{N_t-1}}
{\rm Tr}_{c} \{\hat U_4 (\hat U_4-\hat U_{-4})^{N_t-1}\} 
=\frac{4}{(2a)^{N_t-1}}{\rm Tr}_{c} \{ \hat U_4^{N_t} \}
=-\frac{4N_cV}{(2a)^{N_t-1}}L_P. 
\label{eq:FTdetail}
\end{eqnarray}
Here, a minus appears from Eq.(\ref{eq:PL}), which 
reflects the temporal anti-periodicity of $\Slash D$.
We thus obtain the relation between 
$I= {\rm Tr}_{c,\gamma}
 (\hat U_4 \hat{\Slash D}^{N_t-1})$ 
and the Polyakov loop $L_P$,
\begin{eqnarray}
I={\rm Tr}_{c,\gamma} (\hat U_4 \hat{\Slash D}^{N_t-1}) 
=-\frac{4N_cV}{(2a)^{N_t-1}}L_P.
\label{eq:FTtoPL}
\end{eqnarray}

On the other hand, we calculate the functional trace 
in Eq.(\ref{eq:FTV}) using the complete set of 
the Dirac-mode basis $|n\rangle$ satisfying $\sum_n |n\rangle \langle n|=1$, 
and find the Dirac-mode representation of 
\begin{eqnarray}
I=\sum_n\langle n|\hat{U}_4\Slash{\hat{D}}^{N_t-1}|n\rangle
=i^{N_t-1}\sum_n\lambda_n^{N_t-1}\langle n|\hat{U}_4| n \rangle. 
\label{eq:FTtoD}
\end{eqnarray}
Combing Eqs.(\ref{eq:FTtoPL}) and (\ref{eq:FTtoD}), we obtain the analytical 
formula between the Polyakov loop $L_P$ 
and the Dirac eigenvalues $i\lambda_n$: 
\begin{eqnarray}
L_P =-\frac{(2ai)^{N_t-1}}{4N_cV}
\sum_n\lambda_n^{N_t-1}\langle n|\hat{U}_4| n \rangle
\label{eq:PLvsD}
\end{eqnarray}
for each gauge configuration.
Taking the gauge-configuration average, we obtain
\begin{eqnarray}
\langle L_P \rangle =-\frac{(2ai)^{N_t-1}}{4N_cV}
\left\langle \sum_n\lambda_n^{N_t-1}\langle n|\hat{U}_4| n \rangle 
\right\rangle_{\rm gauge~ave.}
\label{eq:PLvsDave}
\end{eqnarray}
This is a direct relation between the Polyakov loop $\langle L_P\rangle$ 
and the Dirac modes in QCD, and is 
mathematically valid in lattice QCD with odd-number temporal size 
in both confinement and deconfinement phases. 
The formula (\ref{eq:PLvsD}) is 
a Dirac spectral representation of the Polyakov loop, 
and we can investigate each Dirac-mode contribution 
to the Polyakov loop individually, based on Eq.(\ref{eq:PLvsD}). 
(For example, each contribution specified by $n$ 
is numerically calculable in lattice QCD \cite{DSI14}.)
%

As a remarkable fact, because of the factor $\lambda_n^{N_t -1}$, 
the contribution from 
low-lying Dirac-modes with $|\lambda_n|\simeq 0$ 
is negligibly small in the Dirac spectral sum of RHS in Eq.(\ref{eq:PLvsD}),
in comparison with the other Dirac-mode contribution. 
In fact, the low-lying Dirac modes have fairly small contribution 
to the Polyakov loop in Eq.(\ref{eq:PLvsD}), 
regardless of confinement or deconfinement phase.

This is consistent with the previous numerical 
lattice QCD result that confinement properties are almost unchanged by 
removing low-lying Dirac modes from the QCD vacuum \cite{S111213,GIS12,IS14}.


\section{Discussions on the Dirac spectral representation of the Polyakov loop}

In this section, we consider 
the Dirac spectral representation of the Polyakov loop, i.e., 
the formula (\ref{eq:PLvsD}) between the Polyakov loop and Dirac modes, 
and discuss its physical meaning.
In particular, we consider the contribution from low-lying Dirac modes 
to the Polyakov loop. 

\subsection{Properties of the formula between Polyakov loop and Dirac modes}

First, we note that 
Eq.(\ref{eq:PLvsD}) is a manifestly gauge-invariant formula. 
Actually, the matrix element $\langle n |\hat U_4|n\rangle$ 
can be expressed with 
the Dirac eigenfunction $\psi_n(s)$ and 
the temporal link-variable $U_4(s)$ as 
\begin{eqnarray}
\langle n |\hat U_4|n\rangle =
\sum_s \langle n |s \rangle \langle s 
|\hat U_4| s+\hat t \rangle \langle s+\hat t|n\rangle 
=\sum_s \psi_n^\dagger (s)U_4(s) \psi_n(s+\hat t),
\end{eqnarray}
and each term $\psi_n^\dagger (s)U_4(s) \psi_n(s+\hat t)$ 
is manifestly gauge invariant, because of 
the gauge transformation property (\ref{eq:GTDwf}).
Here, the irrelevant global phase factors 
also cancel exactly as $e^{-i\varphi_n}e^{i\varphi_n}=1$ 
between $\langle n|$ and $|n \rangle$ \cite{S111213,GIS12,IS14}.

Second, we note the chiral property and nontriviality of Eq.(\ref{eq:PLvsD}).
In RHS of Eq.(\ref{eq:PLvsD}), 
there is no cancellation between chiral-pair Dirac eigen-states, 
$|n \rangle$ and $\gamma_5|n \rangle$, 
because $(N_t-1)$ is even, i.e., 
$(-\lambda_n)^{N_t-1}=\lambda_n^{N_t-1}$, and  
$\langle n |\gamma_5 \hat U_4 \gamma_5|n\rangle
=\langle n |\hat U_4|n\rangle$. 

Third, Eq.(\ref{eq:PLvsD}) is correct for any odd number $N_t ( > 1)$ 
and is applicable to both confinement and deconfinement phases.
Then, Eq.(\ref{eq:PLvsD}) obtained on the odd-number lattice 
is expected to hold in the continuum limit of $a \rightarrow 0$ 
and $N_t \rightarrow \infty$, since 
any number of large $N_t$ gives the same physical result.

Finally, we comment on generality and 
wide applicability of Eq.(\ref{eq:PLvsD}).
In the argument to derive Eq.(\ref{eq:PLvsD}), 
we only use a few setup conditions:
\begin{itemize}
\item[i)] square lattice (including anisotropic cases)
\item[ii)] odd-number temporal size $N_t (< N_s)$
\item[iii)] temporal periodicity for link-variables.
\end{itemize}
Accordingly, Eq.(\ref{eq:PLvsD}) is widely correct 
in the case of {\it arbitrary gauge group of the theory.} 
For example, Eq.(\ref{eq:PLvsD}) is applicable 
in the SU($N_c$) gauge theory 
for the arbitrary color number $N_c$.
In addition, regardless of presence or absence of dynamical quarks, 
Eq.(\ref{eq:PLvsD}) is formally correct 
as the Dirac-mode expansion. 
In fact, Eq.(\ref{eq:PLvsD}) can be derived also for 
the gauge configuration after integrating out 
quark degrees of freedom.
Of course, the dynamical quark effect appears 
in the Polyakov loop $L_P$, 
the Dirac eigenvalue distribution $\rho(\lambda)$ and 
$\langle n |\hat U_4|n\rangle$.
However, the formula (\ref{eq:PLvsD}) holds 
even in the presence of dynamical quarks. 
Therefore, the formula (\ref{eq:PLvsD}) is applicable 
at finite density and finite temperature.

\subsection{On the small contribution from low-lying Dirac modes 
to the Polyakov loop}

In this subsection, we consider the contribution from 
low-lying Dirac modes to the Polyakov loop based on Eq.(\ref{eq:PLvsD}).
Due to the factor $\lambda_n^{N_t -1}$, the contribution from 
low-lying Dirac-modes with $|\lambda_n|\simeq 0$ 
is negligibly small in RHS in Eq.(\ref{eq:PLvsD}),
compared with the other Dirac-mode contribution, so that 
the low-lying Dirac modes have small contribution 
to the Polyakov loop in both confinement and deconfinement phases.

If RHS in Eq.(\ref{eq:PLvsD}) {\it were} not a sum but a product, 
low-lying Dirac modes, or the small $|\lambda_n|$ region,
should have given an important contribution 
to the Polyakov loop as a crucial reduction factor of $\lambda_n^{N_t-1}$. 
In the sum, however, the contribution ($\propto \lambda_n^{N_t-1}$) 
from the small $|\lambda_n|$ region is negligible. 

Even if $\langle n |\hat U_4|n\rangle$ behaves as the $\delta$-function 
$\delta(\lambda)$, the factor $\lambda_n^{N_t-1}$ is still crucial 
in RHS of Eq.(\ref{eq:PLvsD}), 
because of $\lambda \delta(\lambda)=0$. 
In fact, without appearance of extra counter factor 
$\lambda_n^{-(N_t-1)}$ from $\langle n |\hat U_4|n\rangle$, 
the crucial factor $\lambda_n^{N_t-1}$ inevitably leads to small contribution 
for low-lying Dirac modes.
Note here that the explicit $N_t$-dependence appears 
as the factor $\lambda_n^{N_t-1}$ in RHS of Eq.(\ref{eq:PLvsD}), 
and the matrix element $\langle n |\hat U_4|n\rangle$ 
does not include $N_t$-dependence in an explicit manner.
Then, it seems rather difficult to consider the appearance of 
the counter factor $\lambda_n^{-(N_t-1)}$ 
from the matrix element $\langle n |\hat U_4|n\rangle$.

One may suspect the necessity of renormalization for the Polyakov loop, 
although the Polyakov loop is at present 
one of the typical order parameters of confinement, 
and most arguments on the QCD phase transition have been done 
in terms of the simple Polyakov loop.
Even in the presence of a possible 
multiplicative renormalization factor for the Polyakov loop like $Z_P L_P$,
the contribution from the low-lying Dirac modes, 
or the small $|\lambda_n|$ region, is relatively negligible 
compared with other Dirac-mode contribution 
in the sum of RHS in Eq.(\ref{eq:PLvsD}).

\subsection{Numerical confirmation with lattice QCD}

It is notable that 
all the above arguments can be numerically confirmed by 
lattice QCD calculations. 
In this subsection, we briefly mention 
the numerical confirmation with lattice QCD Monte Carlo calculations 
\cite{DSI14}.

Using actual lattice QCD calculations at the quenched level, 
we numerically confirm the analytical formula (\ref{eq:PLvsD}), 
non-zero finiteness of $\langle n|\hat U_4|n\rangle$ for each Dirac mode, 
and negligibly small contribution of 
low-lying Dirac modes to the Polyakov loop, 
i.e.,  
the Polyakov loop is almost unchanged 
even by removing low-lying Dirac-mode contribution from 
the QCD vacuum generated by lattice QCD simulations, 
in both confinement and deconfinement phases \cite{DSI14}. 

As for the matrix element $\langle n|\hat U_4|n \rangle$, 
its behavior is different between confinement and deconfinement phases.
In the confinement phase, we find a ``positive/negative symmetry'' on 
the distribution of the matrix element $\langle n|\hat U_4|n \rangle$ 
\cite{DSI14}, i.e., its actual value seems to appear as ``pair-wise'' 
of plus and minus, and this symmetry is one of the essence to realize 
the zero value of the Polyakov loop $L_P$. 
In fact, due to this symmetry of $\langle n|\hat U_4|n \rangle$ 
in the confinement phase, 
the contribution from partial Dirac modes in arbitrary region 
$a \le \lambda_n \le b$ leads to $L_P=0$. 
In particular, the high-lying Dirac modes do not contribute to 
the Polyakov loop $L_P$, in spite of the large factor $\lambda_n^{N_t-1}$.
This behavior is consistent with our previous 
lattice QCD results \cite{S111213,GIS12,IS14}, 
which indicate that the ``seed'' of confinement is distributed 
in a wider region of the Dirac eigenmodes unlike chiral symmetry breaking. 
In the deconfinement phase, there is no such 
symmetry on the distribution of $\langle n|\hat U_4|n \rangle$, 
and this asymmetry leads to a non-zero value of the Polyakov loop 
\cite{DSI14}.

In any case, regardless of the behavior of $\langle n|\hat U_4|n \rangle$, 
we numerically confirm that the contribution from 
low-lying Dirac modes to the Polyakov loop is negligibly small \cite{DSI14}
in both confinement and deconfinement phases, 
owing to the factor $\lambda_n^{N_t-1}$ in Eq.(\ref{eq:PLvsD}). 

From the analytical formula (\ref{eq:PLvsD}) and the numerical confirmation, 
low-lying Dirac-modes have small contribution to the Polyakov loop, 
and are not essential for confinement, 
while these modes are essential for chiral symmetry breaking.

\section{Similar formula for the Wilson loop on arbitrary square lattices}

In this section, we attempt a similar consideration to 
the Wilson loop and the string tension on arbitrary square lattices 
(including anisotropic cases) 
with any number of $N_t$, i.e., without restriction of odd-number size.
We consider the ordinary Wilson loop on a $R \times T$ rectangle, 
where $T$ and $R$ are arbitrary positive integers.
The Wilson loop is expressed by the functional trace \cite{S111213,GIS12}
\begin{eqnarray}
W \equiv {\rm Tr}_{c}  \hat U_1^R \hat U_{-4}^T \hat U_{-1}^R \hat U_4^T
={\rm Tr}_{c} \hat U_{\rm staple} \hat U_4^T,
\end{eqnarray}
where we introduce the ``staple operator'' $\hat U_{\rm staple}$ as 
\begin{eqnarray}
\hat U_{\rm staple} \equiv \hat U_1^R \hat U_{-4}^T \hat U_{-1}^R.
\end{eqnarray}
Here, the Wilson-loop operator is factorized as 
a product of $\hat U_{\rm staple}$ and $\hat U_4^T$, 
as shown in Fig.4.
We note that $W \propto \langle W \rangle_{\rm gauge~ave.}$ 
for enough large volume lattice \cite{S111213,GIS12}.

\begin{figure}[h]
\begin{center}
\includegraphics[scale=0.6]{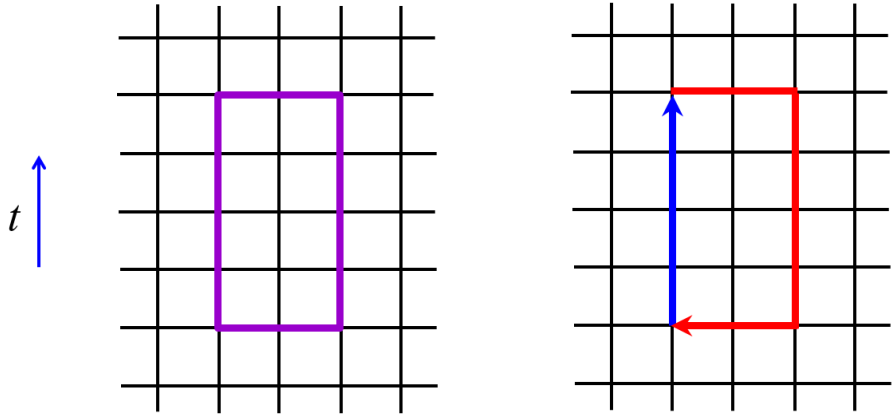}
\caption{
The left figure shows the Wilson loop $W$ defined on a $R \times T$ rectangle. 
The right figure shows the factorization of the Wilson-loop operator 
as a product of 
$\hat U_{\rm staple}\equiv \hat U_1^R \hat U_{-4}^T \hat U_{-1}^R$ 
and $\hat U_4^T$. Here, $T$, $R$, and the lattice size are arbitrary.
}
\end{center}
\end{figure}

In the case of even number $T$, 
let us consider the functional trace of 
\begin{eqnarray}
J \equiv {\rm Tr}_{c,\gamma} \hat U_{\rm staple} \hat {\Slash D}^T.
\end{eqnarray}
From the similar arguments in Sect.3, 
we obtain 
\begin{eqnarray}
J &=& {\rm Tr}_{c,\gamma} \hat U_{\rm staple} \hat {\Slash D}^T 
   = {\rm Tr}_{c,\gamma} \hat U_{\rm staple} (\gamma_4 \hat D_4)^T 
   = 4 {\rm Tr}_{c} \hat U_{\rm staple} \hat D_4^T  \cr
  &=&  \frac{4}{(2a)^T} {\rm Tr}_{c} \hat U_{\rm staple} 
(\hat U_4- \hat U_{-4})^T 
  = \frac{4}{(2a)^T} {\rm Tr}_{c} \hat U_{\rm staple} \hat U_4^T 
  =  \frac{4}{(2a)^T} W, 
\end{eqnarray}
and 
\begin{eqnarray}
J =\sum_{n} \langle n| \hat U_{\rm staple} {\Slash D}^T |n \rangle 
= (-)^{\frac{T}{2}}
\sum_{n} \lambda_n^T \langle n| \hat U_{\rm staple} |n \rangle.~
\end{eqnarray}
Therefore, we obtain for even $T$ the simple formula of 
\begin{eqnarray}
W = \frac {(-)^{\frac{T}{2}}(2a)^T}{4}
\sum_{n} \lambda_n^T \langle n| \hat U_{\rm staple} |n \rangle.
\label{eq:Wilson}
\end{eqnarray}
Again, owing to the factor $\lambda_n^T$, 
the contribution from low-lying Dirac modes is expected to be small 
also for the Wilson loop, although the matrix element 
$\langle n| \hat U_{\rm staple} |n \rangle$ 
includes explicit $T$-dependence 
and its behavior is not so clear, 
unlike the formula (\ref{eq:PLvsD}) for the Polyakov loop.

In the case of odd number $T$, 
the similar results can be obtained 
by considering 
\begin{eqnarray}
K \equiv {\rm Tr}_{c,\gamma} 
\hat U_{\rm staple} \hat U_4 \hat {\Slash D}^{T-1}
\end{eqnarray}
instead of $J$.
Actually, one finds 
\begin{eqnarray}
K &=& {\rm Tr}_{c,\gamma} \hat U_{\rm staple} \hat U_4 \hat {\Slash D}^{T-1}   
   = {\rm Tr}_{c,\gamma} \hat U_{\rm staple} \hat U_4 (\gamma_4 \hat D_4)^{T-1}
   = 4 {\rm Tr}_{c} \hat U_{\rm staple} \hat U_4 \hat D_4^{T-1} \cr
  &=&  \frac{4}{(2a)^{T-1}} {\rm Tr}_{c} \hat U_{\rm staple} \hat U_4
 (\hat U_4- \hat U_{-4})^{T-1} 
   =  \frac{4}{(2a)^{T-1}} {\rm Tr}_{c} \hat U_{\rm staple} \hat U_4^T 
   =  \frac{4}{(2a)^{T-1}} W, 
\end{eqnarray}
and 
\begin{eqnarray}
K =\sum_{n} \langle n| 
\hat U_{\rm staple} \hat U_4 {\Slash D}^{T-1} |n \rangle 
= (-)^{\frac{T-1}{2}}
\sum_{n} \lambda_n^{T-1} \langle n| \hat U_{\rm staple} \hat U_4 |n \rangle,~
\end{eqnarray}
so that one finds for odd $T$ the similar formula of 
\begin{eqnarray}
W = \frac {(-)^{\frac{T-1}{2}}(2a)^{T-1}}{4}
\sum_{n} \lambda_n^{T-1} \langle n| \hat U_{\rm staple} \hat U_4|n \rangle.
\label{eq:Wilsonodd}
\end{eqnarray}

Finally, for even $T$ case, we show the inter-quark potential $V(R)$ 
and the string tension $\sigma$. 
From the expression (\ref{eq:Wilson}) for the Wilson loop $W$, 
we obtain the inter-quark potential $V(R)$ 
and the string tension $\sigma$: 
\begin{eqnarray}
V(R) &=&-\lim_{T \to \infty} 
\frac{1}{T}{\rm ln} W 
= -\lim_{T \to \infty}\frac{1}{T}
{\rm ln} \left|\sum_{n} 
(2a \lambda_n)^T \langle n| \hat U_{\rm staple} |n \rangle\right|,~~ 
\end{eqnarray}
\begin{eqnarray}
\sigma &=&-\lim_{R,T \to \infty} \frac{1}{RT}{\rm ln} W 
= -\lim_{R,T \to \infty}\frac{1}{RT}
{\rm ln} \left|\sum_{n} 
(2a \lambda_n)^T \langle n| \hat U_{\rm staple} |n \rangle\right|.~~
\end{eqnarray}
Because of the factor $\lambda_n^T$ in the sum, 
the low-lying Dirac-mode contribution is to be small for the Wilson loop $W$, 
the inter-quark potential $V(R)$ and the string tension $\sigma$,
unless the extra counter factor $\lambda_n^{-T}$ appears from 
$\langle n |\hat U_{\rm staple}|n\rangle$.
Also for odd $T$ case, similar arguments can be done 
with Eq.(\ref{eq:Wilsonodd}).

In this way, the string tension $\sigma$, or the confining force, 
is expected to be unchanged by the removal of 
the low-lying Dirac-mode contribution, 
which is consistent with our previous 
numerical works of lattice QCD \cite{S111213,GIS12}.

Taking the assumption (\ref{eq:assumption}) in Sec.2.4, 
the removal of low-lying Dirac modes of $\Slash D$ leads to 
an approximate removal of low-lying modes of the fermionic kernel $K[U]$, 
which largely reduces the chiral condensate.
%
Then, the above suggested insensitivity of confinement 
to low-lying Dirac modes 
indicates no direct one-to-one correspondence 
between confinement and chiral symmetry breaking in QCD.

\section{Summary and Concluding Remarks} 

We have derived 
an analytical gauge-invariant formula between the Polyakov loop 
$L_P$ and the Dirac eigenvalues $\lambda_n$ as 
$L_P \propto 
\sum_n \lambda_n^{N_t -1} \langle n|\hat U_4|n \rangle$ 
in lattice QCD with odd-number temporal size $N_t$,
by considering ${\rm Tr} (\hat{U}_4\hat{\Slash{D}}^{N_t-1})$, 
on the ordinary square lattice with 
the normal (nontwisted) temporally-periodic boundary condition 
for link-variables.
Here, $|n\rangle$ denotes the Dirac eigenstate, and $\hat U_4$ 
temporal link-variable operator.

This formula is a Dirac spectral representation of 
the Polyakov loop in terms of Dirac eigenmodes $|n\rangle$, 
and expresses 
each contribution of the Dirac eigenmode to the Polyakov loop.
Because of the factor $\lambda_n^{N_t -1}$ in the Dirac spectral sum, 
this formula indicates fairly small contribution of 
low-lying Dirac modes to the Polyakov loop 
in both confinement and deconfinement phases, 
while these modes are essential for chiral symmetry breaking. 

Next, we have found a similar formula between 
the Wilson loop and Dirac modes on arbitrary lattices, 
without restriction of odd-number size. 
This formula suggests a small contribution of low-lying Dirac modes 
to the string tension $\sigma$, or the confining force.

Thus, it is likely that low-lying Dirac-modes have fairly small contribution 
to the Polyakov loop and the string tension, 
and are not essential modes for confinement, 
while these modes are essential for chiral symmetry breaking. 
This suggests no direct one-to-one 
correspondence between confinement and chiral symmetry breaking in QCD.
Note here that the independence of confinement and chiral symmetry breaking 
would be natural, because heavy quarks are also confined 
even without light quarks or the chiral symmetry. 

Also for thermal QCD, we have investigated the relation between 
confinement and chiral symmetry breaking 
using the ratio of the susceptibility of the Polyakov loop \cite{DRSS15}, 
of which importance for the deconfinement transition 
has been pointed out \cite{LFKRS13}. 

Finally, we state some cautions and future works in this framework in order. 

In Sec.2.4, for the strict connection to chiral symmetry breaking, 
we have assumed that each low-lying mode of the fermionic kernel $K[U]$ 
is mainly expressed with the low-lying Dirac modes of $\Slash D$. 
We are now investigating this assumption (\ref{eq:assumption}) 
quantitatively in lattice QCD \cite{DS15}.

In this paper, we have derived the Dirac-mode expansion such as 
Eq.(\ref{eq:PLvsD}), which is mathematically correct. 
In this expansion, each Dirac-mode contribution is explicitly expressed, 
and we have focused on this explicit contribution. This treatment 
would be appropriate in quenched QCD.
For more definite argument in full QCD, however, we have to clarify 
an implicit contribution of the Dirac modes in the fermionic determinant, 
which can actually alter the properties of the QCD vacuum \cite{H09}.

It is important to take the continuum limit 
of the mathematical formulae obtained on the lattice, 
although it seems a difficult problem.
It is also interesting to compare with other lattice QCD result 
on the important role of infrared gluons (below about 1GeV) 
for confinement in the Landau gauge \cite{YS0809}, 
in contrast to the insensitivity of confinement 
against low-lying Dirac-modes.

This work suggests some possible independence between confinement 
and chiral symmetry breaking in QCD, and this may lead to 
richer phase structure of QCD in various environment.
In fact, there is an interesting possibility that 
QCD phase transition points can be generally different 
between deconfinement and chiral symmetry restoration, 
e.g., in the presence of strong electromagnetic fields, 
because of their nontrivial effect on the chiral symmetry \cite{ST91}.

\section*{Acknowledgements}
H.S. thanks Prof. E.T. Tomboulis for useful discussions 
and his brief confirmation on our calculations.
H.S. is supported in part by the Grant for Scientific Research 
[(C) No.23540306, 15K05076, E01:21105006] 
from the Ministry of Education, Science and Technology of Japan.

\end{document}